\documentclass[aps,twocolumn,pra,superscriptaddress,showpacs]{revtex4}

\usepackage{mathrsfs}
\usepackage{amssymb}
\usepackage{amsmath}
\usepackage{graphicx}
\usepackage{epsfig}
\usepackage{txfonts}
\usepackage{subfigure}
\usepackage{amsfonts}
\usepackage{color}
\usepackage[colorlinks,citecolor=blue]{hyperref}
\usepackage{bm}

\begin{document}
	
	\title{Enhancing the sensitivity of quantum optomechanical gyroscope by optical Kerr effect}
	
	\author{Ying Liu}\thanks{Co-first authors with equal contribution}
	\affiliation{Key Laboratory of Low-Dimensional Quantum Structures and Quantum Control of Ministry of Education, Synergetic Innovation Center for Quantum Effects and Applications,  XJ-Laboratory and Department of Physics, Hunan Normal University, Changsha 410081, China}
	
	\author{Rui Zhang}\thanks{Co-first authors with equal contribution}

	\affiliation{Key Laboratory of Low-Dimensional Quantum Structures and Quantum Control of Ministry of Education, Synergetic Innovation Center for Quantum Effects and Applications,  XJ-Laboratory and Department of Physics, Hunan Normal University, Changsha 410081, China}
	
\author{Wen-Quan Yang}
	\affiliation{Key Laboratory of Low-Dimensional Quantum Structures and Quantum Control of Ministry of Education, Synergetic Innovation Center for Quantum Effects and Applications,  XJ-Laboratory and Department of Physics, Hunan Normal University, Changsha 410081, China}

\author{Ya-Feng Jiao}\email{yfjiao@zzuli.edu.cn}
	\affiliation{Academy for Quantum Science and Technology, Zhengzhou University of Light Industry, Zhengzhou 450002, China}

\author{Wang-Jun Lu}\email{wjlu1227@zju.edu.cn}
\affiliation{Department of Maths and Physics, Hunan Institute of Engineering, Xiangtan 411104, China}

\author{Qing-Shou Tan}  \email{qstan@hnist.edu.cn}
	\affiliation{Keg Laboratory of Hunan Province on Information Photonics and Freespace Optical Communication,
College of Phgsics and Electronics, Hunan Institute of Science and Technology, Yueyang, Hunan 414000, China}	
	
	\author{Le-Man Kuang}\email{lmkuang@hunnu.edu.cn}
	\affiliation{Key Laboratory of Low-Dimensional Quantum Structures and Quantum Control of Ministry of Education, Synergetic Innovation Center for Quantum Effects and Applications,  XJ-Laboratory and Department of Physics, Hunan Normal University, Changsha 410081, China}
	\affiliation{Academy for Quantum Science and Technology, Zhengzhou University of Light Industry, Zhengzhou 450002, China}
	


\begin{abstract}
		
We propose  a theoretical scheme to enhance the sensitivity of a quantum optomechanical gyroscope (QOMG)  by optical Kerr effect. We utilize quantum Fisher information (QFI) to evaluate the
metrological potential of the QOMG scheme. It is found that  the Kerr interaction can significantly enhances the sensitivity of the QOMG. We observe the super-Hesenberg scaling of parameter estimation precision.  Furthermore, we also evaluate the performance of QOMG for the quadrature measurement. It is indicated that the sensitivity in the quadrature measurement scheme can saturate the quantum Crmam\'{e}r-Rao bound. We study the effect of the driving and dissipation of the optical cavity on  the QFI, and find that  the sensitivity can be manipulated by changing the driving while  dissipation decreases the sensitivity. The work shows that the  photon nonlinear interaction can improve sensitivity of QOMG, and demonstrates a valuable quantum resource for the QOMG. These results could have a wide-ranging impact on developing  high-performance QOMG in the future.
		
\end{abstract}

  \maketitle
\section{\label{level}Introduction}

Cavity optomechanical systems  \cite{Aspelmeyer,Bowen,Kippenberg,Huang} have become an important platform both for fundamental physics of macroscopic quantum systems  \cite{Lepinay,Wollman,Purdy,Chegnizadeh,Kiesewetter,Jiao1,Jiao2} and for practical applications of precision sensing \cite{LiBB,Forstner,Westerveld,Arcizet}.
The cavity fields in cavity optomechanical systems exert radiation pressure on the movable mirror, which leads to the changes of both the resonance frequency and damping rate of the mechanical modes. At the same time, the mechanical vibration of the spring modulates the position of the movable mirror, which changes the cavity length and optical resonant frequency.
The resonant enhancement of both mechanical and optical response in the cavity optomechanical systems has enabled precision sensing of multiple physical quantities, including displacements, masses, forces, accelerations, magnetic fields, and ultrasounds.
Based on the ultrasensitive displacement measurement, precision sensing of various physical quantities, such as force and mass, etc., has been realized. Besides the ultrahigh precision, cavity optomechanical sensors also provide the advantages of small size, low weight, and low power consumption, on-chip integration capability, compatibility with fiber coupling, etc., and therefore have great potential to be used in real applications in the near future.

The gyroscope is a device to measure  angular rotation around a fixed axis with respect to an inertial space. It is a key sensor in modern navigation systems enabling to plan, record and control the movement of a vehicle from one place to another. This device has wide applications in space engineering, aeronautical and military industry, automotive, medicine and so on.
Quantum  optomechanical gyroscope (QOMG) is a kind of quantum gyroscope based on cavity optomechanical systems. It utilizes quantum resources such as quantum squeezing and quantum coherence to enhance the sensitivity. The QOMG aims to realize rotation sensing with high precision by using quantum properties of cavity optomechanical systems.

In recent years, several QOMG schemes have been proposed. Davuluri \cite{Davuluri1} present an application of optomechanical cavity for the absolute rotation detection. The optomechanical cavity is arranged in a Michelson interferometer in such a way that the centrifugal force due to rotation changes the length of the optomechanical cavity. The change in the cavity length induces a shift in the frequency of the cavity mode. The phase shift corresponding to the frequency shift is measured at the interferometer output to estimate the angular velocity of absolute rotation.
Then, Davuluri and coworkers \cite{Davuluri2}  proposed an application of two-dimensional optomechanical oscillator as a gyroscope by detecting the Coriolis force which is modulated at the natural frequency of the optomechanical oscillator.  Li \emph{et al.} \cite{Likai1} developed a coherent quantum noise cancellation (CQNC) method  to beat standard quantum limit  for improving the performance of QOMG. The protocol for realizing CQNC is achieved by constructing an effective negative mass mechanical oscillator simulated by an ancillary cavity. This oscillator shows an anti-response relative to that of a real mechanical oscillator. Thus, the optomechanical back-action noise is counteracted or restrained.
They also proposed a scheme for measuring the angular velocity of absolute rotation using a three-mode optomechanical
system in which one mode of the two-dimensional mechanical resonator is coupled to an optical cavity  \cite{Likai2}.  Mahdavi \emph{et al.} \cite{Mahdavi} presented a design of a quantum optomechanical gyroscope, consisting of a Michelson interferometer
on a rotating table. An optical cavity with two moving mirrors is placed on one of the interferometer arms.  It is shown that by using two moving mirrors, the sensitivity has been improved by more than $56$ percent compared with the cavity with one moving mirror.
Li and Cheng \cite{Lihao} studied the quantum Fisher information (QFI) of the angular velocity of rotation in an optomechanical system.
Based on the Gaussian measurements method, they derived the explicit form of a single-mode Gaussian QFI, which is valid
for arbitrary angular velocity of rotation.  A multiparameter estimation scheme of a QOMG  through two optomechanical subsystems was proposed to simultaneously
estimate the angular velocity and rotation-axis position. Tan et al., proposed  reinforcement learning assisted non-reciprocal QOMG \cite{Tanqs}, which reveals a striking dependence of the gyroscope's sensitivity on the propagation direction of the driving optical field, manifesting robust quantum non-reciprocal behavior.  This non-reciprocity significantly enhances the precision of angular velocity estimation, offering a unique advantage over conventional gyroscopic systems.

Enhancing the precision of measured parameters remains a fundamental topic in quantum metrology and quantum sensing \cite{1Giovannetti,2Giovannetti,3Giovannetti,4Apellaniz,5Barbieri,6Degen,7Pirandola}. Nonclassical effects of quantum systems \cite{8Kwon} can be used to improve the performance of various sensing systems beyond the shot noise limit that applies to classical sources. Quantum-enhanced metrology and sensing study how to exploit quantum resources to enhance the sensitivity of parameter estimations. It has already been demonstrated that squeezed states \cite{9Caves,10LIGO,11Aasi,12LIGO,13Goodwin,14Ganapathy,15Anisimov,16Schnabel,17Zuo,18Zhao,19Pezz,20Nielsen}, entangled states
	\cite{21Zhang,22Wasilewski,23Zhao,24Belliardo,25Hao,26Boto,27Kok,28Joo,29Chen,30Maccone,31Li,32Zhuang,33Lloyd,34tTan,35Gallego,36Giovannetti,37Lee,38Joo,39Dowling}, and quantum phase transitions \cite{40Chen,41Chu,42Zhang,43Chu,44Lu,45Xu,46Jing,47Peng} are important quantum resources that offer quantum advantages in many emerging sensing applications.
On the other hand, it has been shown that interactions among particles   can contribute to measurement sensitivity and give scaling beyond Heisenberg limit, so-called super-Heisenberg scaling \cite{Boixo}. Relevant interactions include  Kerr nonlinearities \cite{Beltran,Napolitano},
  nonlinear interactions in  atomic Bose-Einstein condensates \cite{Choi} and  atomic ensembles \cite{Chase}, mechanical nonlinearity in nanomechanical resonators \cite{Woolley}, and so on.

In this paper, we propose  a theoretical scheme to enhance the sensitivity of QOMG by the optical Kerr effect.  Our QOMG scheme is employed by using a optical cavity which is filled with  nonlinear Kerr medium.  We investigate the effects of the Kerr nonlinearity and external single-photon driving on sensing performance.
We estimate the metrological potential of the resulting Kerr-QOMG using theoretical verification such as quantum Fisher information (QFI), and then
we observe that increasing the controlling parameters such as Kerr nonlinearity and changing driving field significantly improves the estimation precision of QOMG.
To enhance our analysis, we specifically evaluate the performance of a practical measurement scheme, the quadrature measurement, demonstrate that the quadrature measurement is optimal.

This paper is structured as follows. In Sec. II, we present QOMG model with optical Kerr interaction and derive the Hamiltonian of the QOMG model. In Sec. III, we study the sensitivity of the QOMG by analyzing the QFI  when the optical and mechanical mode is in a product state of  two coherent states in the absence of the driving field. In Sec. IV,  we investigate the classical Fisher information of the QOMG for the quadrature measurement and compare it with the QFI. In Sec. V, we discuss the effect of the driving and dissipation of the cavity on the performance of the QOMG.  Finally, concluding remarks are summarized in Sec. VI.
	
	\begin{figure}[t]
		\centerline{
			\includegraphics[width=0.48\textwidth]{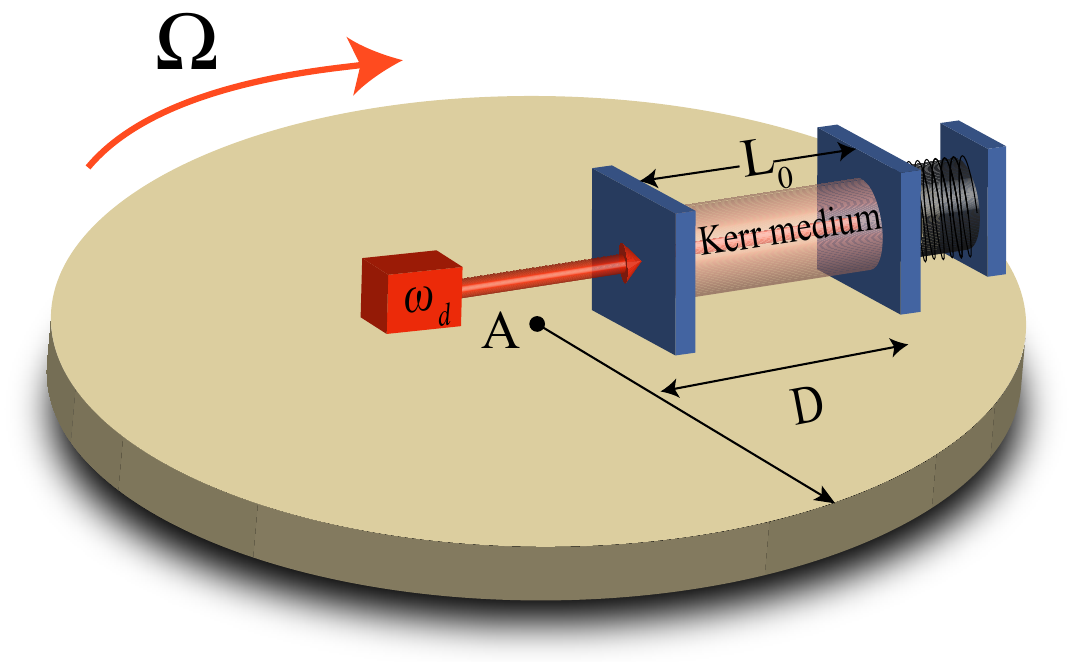}}
		\caption{Schematic of the quantum optomechanical gyroscope with optical Kerr interaction.} \label{fig:1}
	\end{figure}

	\section{\label{level2}  QOMG model with Kerr interaction }

We consider a QOMG model with Kerr interaction  as shown in Fig.~\ref{fig:1}.  It consists of a optomechanical cavity filed with Kerr medium with frequency of the cavity and the mechanical  oscillator being $\omega_{c}$ and  $\omega_{m}$, respectively. The cavity is driven by a coherent laser with driving strength $\varepsilon $ and  frequency $\omega_{d}$. The system is fixed on a rotating table with angular velocity $\Omega$.  The cavity mirror is in a fixed position. The mechanical mirror with the mass $m$ connects to one side of a spring while the other side  of the spring is fixed on the table. The mass of the spring is negligible. The radiation pressure force displaces the cavity frequency and the mechanical mirror by a length of $q$ from its equilibrium position $D$  (where $D$ is the distance between the rotation axis and the mechanical oscillator). Using the approach in Refs. \cite{Kumar,Mikkelsen}, the quantized Hamiltonian of the  QOMG model with Kerr interaction  can be written as
\begin{equation}\label{eq:1}
H=H_{\text{G}}+H_{\text{K}} +H_{\text{D}}
\end{equation}
where the first term on the right-hand side of above equation is the Hamiltonian of the QOMG in the absence of the Kerr medium and the driving field
\begin{equation}
\begin{aligned}
H_{\text{G}} & =\hbar \omega_c a^{\dagger} a+\frac{p^2}{2 m}+\frac{1}{2} m \omega_m^2 q^2+\hbar G a^{\dagger} a q \\
& +\frac{1}{2} m(D-q)^2 \Omega^2 .
\end{aligned}
\end{equation}
where $\hbar$ is Planck constant, $a$ ($a^{\dagger}$) is the annihilation (creation) operator of the single-mode optical field in cavity with frequency $\omega_c$.  $q$ and $p$  represent position and momentum operators of the mechanical oscillator with frequency $\omega_m$ and mass $m$. The coupling between the optical field and mechanical oscillator is described by coupling rate $G$.  The optomechanical system rotates about the axis at an angular velocity $\Omega$, generating rotational kinetic energy.

The second term on the right-hand side of Eq.~(\ref{eq:1}) describes the photon-photon Kerr interaction  in the cavity
\begin{equation}
H_{\text{K}}  =\hbar \eta(q) a^{\dagger 2} a^2
\end{equation}
where $\eta=3 \hbar \omega^2 \operatorname{Re}\left[\chi^{(3)}\right] /\left(2 \epsilon_0 V\right)$ is the Kerr nonlinear coefficient which is dependent of the position of the mechanical mirror. Here $\epsilon_0$ is the vacuum permittivity and $V$ is the cavity volume. Assume that when the mechanical oscillator is at its equilibrium position, the length and volume of the cavity is $L_0$ and $V_0$, respectively.  The volume of the cavity becomes  $(L_0-q)V_0/L_0$ when  the mechanical oscillator is at the position $q$, and we assume that $q\ll L_{0}$. Then, the Kerr nonlinear coefficient can be approximately expressed as
\begin{equation}
\eta(q) \approx \eta_0+g_{\mathrm{NL}} q.
\end{equation}
where $\eta_0=3 \hbar \omega^2 \operatorname{Re}\left[\chi^{(3)}\right] /\left(2 \epsilon_0 V_0\right)$, and $\quad g_{\text{NL}}=3 \eta_0 / L_0$.

The third term on the right-hand side of Eq.~(\ref{eq:1}) is a optical drive term with the amplitude $\varepsilon$ and the frequency $\omega_d$ given by
\begin{equation}
H_{\text {D}}=\hbar \varepsilon a^{\dagger} e^{-i \omega_d t}+\hbar \varepsilon^* a e^{i \omega_d t}.
\end{equation}

In the rotating frame of the driving field,  the total Hamiltonian of the QOMG with optical Kerr interaction can be written as
\begin{equation}\label{eq:6}
\begin{aligned}
H= & \hbar\left(\omega_c-\eta_0\right) a^{\dagger} a+\hbar \eta_0\left(a^{\dagger} a\right)^2+\hbar \tilde{\omega}_m b^{\dagger} b \\
& +\hbar\left[\left(g_0-G_{\mathrm{NL}}\right) a^{\dagger} a+G_{\mathrm{NL}}\left(a^{\dagger} a\right)^2-\chi\right]\left(b+b^{\dagger}\right) \\
&+ \hbar \varepsilon a^{\dagger} +\hbar \varepsilon^* a,
\end{aligned}
\end{equation}
where the frequency of mechanical oscillator has been modified as $\tilde{\omega}_m=\sqrt{\omega_m^2+\Omega^2}$, $g_0=G \sqrt{\hbar /\left(2 m \tilde{\omega}_m\right)}$ is the linearity optomechanical coupling constant, $G_{\mathrm{NL}}=g_{\mathrm{NL}} \sqrt{\hbar /\left(2 m \tilde{\omega}_m\right)}$ shows affects of the nonlinear interaction between the cavity mode and the mechanical oscillator and $\chi=D \Omega^2 \sqrt{m /\left(2 \hbar \tilde{\omega}_m\right)}$ reflects the centripetal force. In Eq.~(\ref{eq:6}) we have expressed  the position and momentum operators of the  mechanical  oscillator with  the frequency $\tilde{\omega}_m$ in terms of  the creation and annihilation operators $b$ and $b^{\dagger}$
\begin{equation}
q=\sqrt{\hbar /\left(2 m \tilde{\omega}_m\right)}\left(b+b^{\dagger}\right), \hspace{0.5cm} p=i \sqrt{m \tilde{\omega}_m \hbar / 2}\left(b-b^{\dagger}\right).
\end{equation}

	\section{\label{level2} Quantum Fisher information}
	
In quantum parameter estimation theory \cite{Helstrom1,Helstrom2}, the quantum Fisher information (QFI) associated with  the estimation parameter $\Omega$ gives the low bound of the sensitivity of the estimation parameter, the quantum Cram\'{e}r-Rao bound (QCRB), which is defined by $\Delta\Omega_{\text{min}}=1/\sqrt{\upsilon F_Q}$ with $F_Q$ being the QFI associated with  the estimation parameter  and $\upsilon$ being the number of repeated measurements (hereafter, we set  $\upsilon=1$ ). Hence, the larger the QFI, the higher the precision of  parameter estimation.
In quantum metrology \cite{Escher}, precision is bounded by the standard quantum limit (SQL), $F\propto N$, and the Heisenberg limit (HL), $F\propto N^2$, where $N$ is the total
number of particles or resources. $F/N\gg 1$ implies that the precision of parameter estimation  surpasses the SQL and exhibits the quantum-enhanced advantage for superior sensitivity.

In this section, we study the QFI of the QOMG with Kerr interaction. For the simplicity, we consider the case in the absence of the driving field, i.e., $\varepsilon =0$. In this case, the Hamiltonian~(\ref{eq:6}) is exactly solvable. The QFI of the QOMG can be analytically calculated using the parameter-generator method of a Hamiltonian \cite{Pang}.

For a pure state $|\Psi\rangle$  with the encoding parameter $\Omega$, the QFI with respect to the parameter $\Omega$ is given by
\begin{equation}\label{eq:8}
F_{Q}=4\left[\left\langle\partial_{\Omega} \Psi \mid \partial_{\Omega} \Psi\right\rangle-\left|\left\langle\Psi \mid \partial_{\Omega} \Psi\right\rangle\right|^2\right].
\end{equation}
For a Hamiltonian under consideration $H$,  the parameterized state can be expressed as $\left|\Psi\right\rangle=U|\psi\rangle$  for the initial pure state $|\psi\rangle$, the unitary  transformation is given by $U=e^{-i t H / \hbar}$.  The QFI can be expressed by
\begin{equation}\label{eq:9}
F_{Q}=4\left[\langle\psi| \mathcal{H}^2|\psi\rangle-\langle\psi| \mathcal{H}|\psi\rangle^2\right].
\end{equation}
where  $\mathcal{H}$ is the generator of parameter translation with respect to the parameter $\Omega$. It can be expanded in terms of the Hamiltonian of the system as
\begin{equation}\label{eq:10}
\mathcal{H}=i \sum_{k=0}^{\infty} \frac{(i t / \hbar)^{k+1}}{(k+1)!} H^{\times k}\left(\partial_{\Omega} H\right),
\end{equation}
where the $k$-th order communtator beween $H$ and $\partial_{\Omega}H$  is defined by
\begin{equation}\label{eq:11}
H^{\times k}\left(\partial_{\Omega} H\right)=[H, [H, \cdot\cdot\cdot, [H,\partial_{\Omega} H]]]_k.
\end{equation}

	\begin{figure}[t]
		\centerline{
			\includegraphics[width=0.4\textwidth]{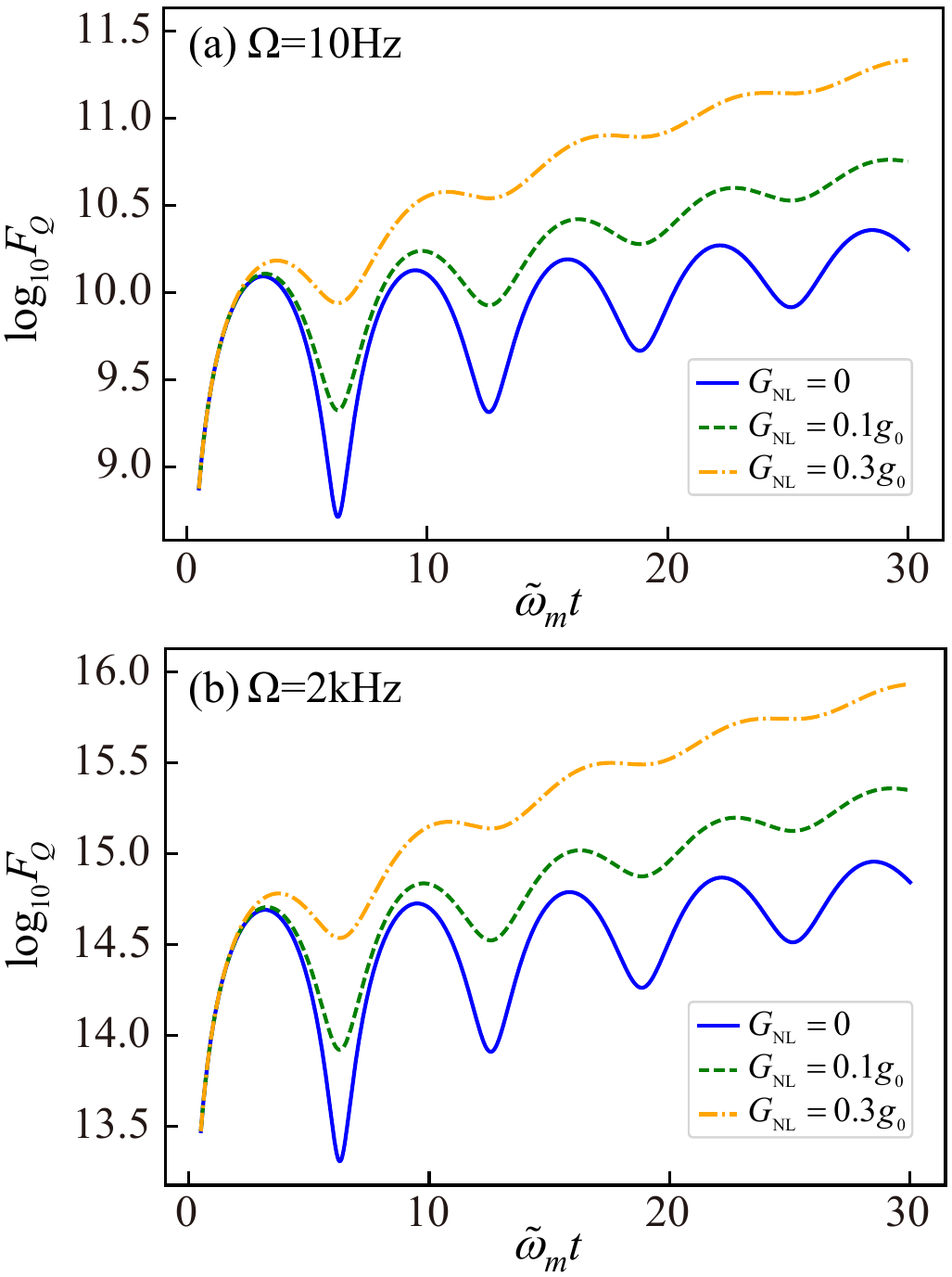}}
		\caption{Dynamic evolution of the QFI for different values of the angular velocity and the nonlinear coupling strength. The parameters of the QOMG system are chosen as $\omega_c=10^{15}$Hz, $\omega_m=62.8$ kHz,  $L_0=10^{-4}$m , $D=10^{-3}$m, $m=10^{-7}$kg. The initial state parameters are taken as $N_C=5$, $N_m=1$.}\label{fig:2}
	\end{figure}
For the Hamiltonian of the QOMG with optical Kerr interaction given by Eq.~(\ref{eq:6}), in the absence of the driving ($\varepsilon =0$), the generator of parameter translation can be expressed as (see Appendix~\ref{appenA} for details)
\begin{equation}\label{eq:12}
\mathcal{H}=-\frac{\Omega}{2 \tilde{\omega}_m^3} \sum_{i=1}^5 \mathcal{H}_i,
\end{equation}
where $\mathcal{H}_i$ is given by Eq.~(\ref{A9}) in Appendix~\ref{appenA}.

From Eqs.~(\ref{eq:9}) and~(\ref{eq:12}), we can express the QFI as
\begin{equation}\label{eq:13}
F_{Q}=\frac{\Omega^2}{\tilde{\omega}_m^6}\left[\sum_{i=1}^5 \operatorname{Var}\left(\mathcal{H}_i\right)+\sum_{j, k=1, j \neq k}^5 \operatorname{Cov}\left(\mathcal{H}_j, \mathcal{H}_k\right)\right]
\end{equation}
where $\operatorname{Var}(\mathcal{H}_i)=\left\langle \mathcal{H}_i^2\right\rangle-\langle \mathcal{H}_i\rangle^2$ is the variance of $\mathcal{H}_i$, and $\operatorname{Cov}\left(\mathcal{H}_i, \mathcal{H}_j\right)=\frac{1}{2}\left\langle\left\{\mathcal{H}_i, \mathcal{H}_j\right\}\right\rangle-\left\langle \mathcal{H}_i\right\rangle\langle \mathcal{H}_j\rangle$ is the covariance. We assume that the optical field in the cavity and the mechanical oscillator are initially in coherent states, $|\alpha\rangle$ and $|\beta\rangle$, respectively. Then the number of photons in the cavity and the number of phonons of the mechnical oscillator are $N_c=|\alpha|^2$ and $N_m=|\beta|^2$, respectively. In this case, the variances and covariance matrix elements on $\mathcal{H}_i$ are explicitly calculated in the Appendix~\ref{appenA}.

In what follows, we numerically investigate the effect of Kerr nonlinear interaction on the QFI to explore strategies of effectively enhancing the sensitivity of the QOMG. Firstly, we study quantum dynamics of the QFI during the time evolution of the QOMG.  In Fig.~\ref{fig:2}, we plot dynamic evolution of the QFI for different values of the angular velocity and the nonlinear coupling strength.The system's parameters of the QOMG are chosen as $\omega_c=10^{15}$Hz, $\omega_m=62.8$ kHz,  $L_0=10^{-4}$m , $D=10^{-3}$m, $m=10^{-7}$kg. The initial state parameters are taken as $N_C=5$, $N_m=1$. Fig.~\ref{fig:2} shows that the strength of the Kerr interaction can significantly enhances the QFI. This implies that the sensitivity of the QOMG can be pronounced improved by the optical Kerr interaction. From Fig.~\ref{fig:2}(a) and Fig.~\ref{fig:2}(b), we can observe that the longer the evolution time, the larger the QFI becomes. And we note that the QFI in the presence of Kerr interaction is two orders of magnitude  higher than that in the absence of Kerr interaction. This means that the sensitivity has more than one order improvement over the situation in the absence of Kerr interaction.
We can also see that at the same time, the higher the angular velocity, the higher the sensitivity of the QOMG.

\begin{figure}[t]
		\centerline{
			\includegraphics[width=0.4\textwidth]{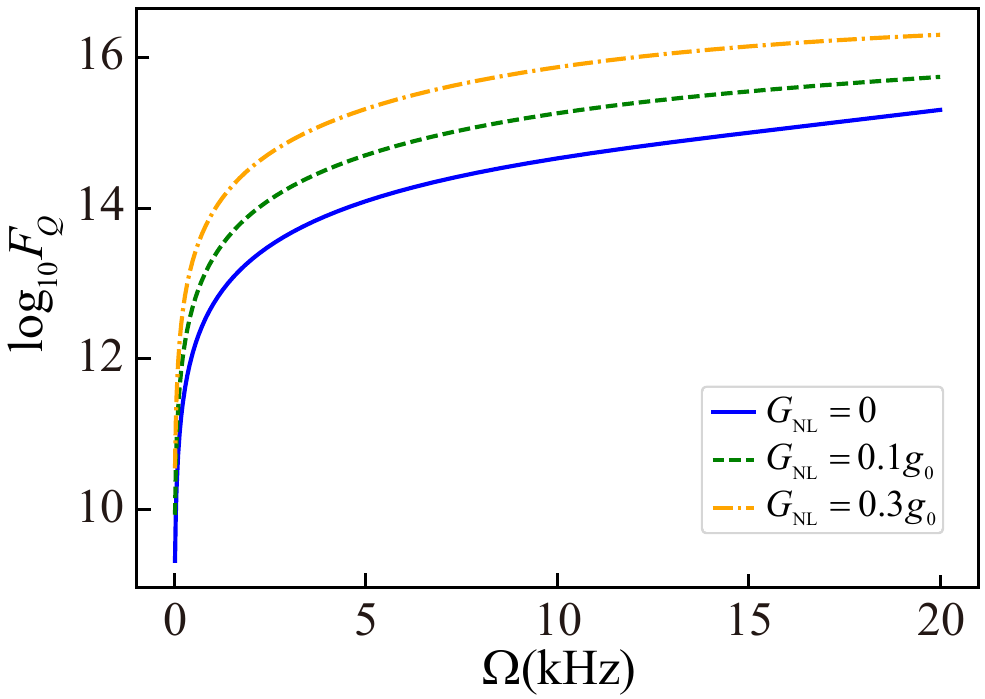}}
		\caption{The QFI with respect to the angular velocity of the QOMG.  The   parameters of the QOMG system are chosen as $\omega_c=10^{15}$Hz, $\omega_m=62.8$ kHz,  $L_0=10^{-4}$m , $D=10^{-3}$m, $m=10^{-7}$kg. The initial state parameters are taken as $N_C=5$, $N_m=1$. The evolution time obeys $\tilde{\omega}_m t=2\pi$.} \label{fig:3}
	\end{figure}

In order to understand the working range of the QOMG, in Fig.~\ref{fig:3} we plot the QFI with respect to the angular velocity of the QOMG for different values of the angular velocity and the nonlinear coupling strength. Here,  the system's parameters of the QOMG are chosen as $\omega_c=10^{15}$Hz, $\omega_m=62.8$ kHz,  $L_0=10^{-4}$m , $D=10^{-3}$m, $m=10^{-7}$kg. The initial state parameters are taken as $N_C=5$, $N_m=1$. The evolution of the system time is chosen as $\tilde{\omega}_m t=2\pi$. Fig.~\ref{fig:3}  indicates that the Kerr interaction can always have the capability to enhance  the QFI in the whole range of the angular velocity. In general, the QFI is enhanced  with  increasing the strength of the Kerr interaction.

\begin{figure}[t]
		\centerline{
			\includegraphics[width=0.4\textwidth]{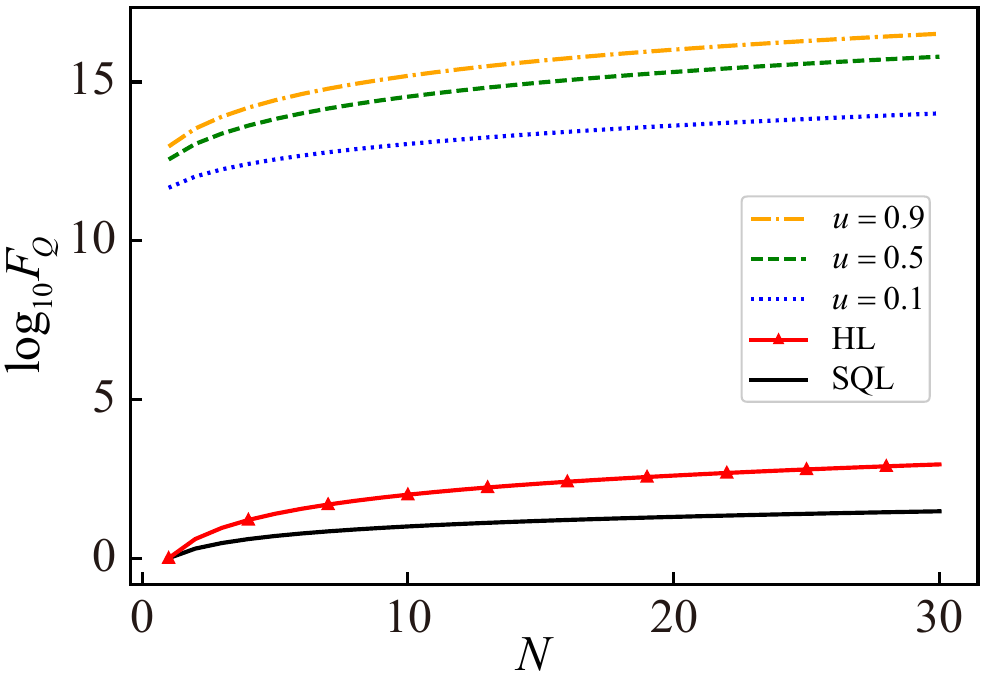}}
		\caption{The QFI with respect to  total number of particles in the input state of the QOMG. Here we take the angular velocity $\Omega=2$kHz and the evolution time $\tilde{\omega}_m t=2\pi$. The parameters of the QOMG system are chosen as $G_{\text{NL}}=0.3g_0$, $\omega_c=10^{15}$Hz, $\omega_m=62.8$ kHz,  $L_0=10^{-4}$m , $D=10^{-3}$m, $m=10^{-7}$kg.} \label{fig:4}
	\end{figure}

We now turn to explore the QFI how to change with the total resources of the QOMG. In Fig.~\ref{fig:4}, we plot the QFI with respect to  total number of particles in the input state of the QOMG when the angular velocity takes $\Omega=2$kHz. Assume that the initial state of the QOMG is initially in the coherent state $|\alpha\rangle|\beta\rangle$.  The total number of  photons and phonons in the initial state is $N=|\alpha|^2 + |\beta|^2$.  In general, under the condition of the same total number of particles $N$, different population of  photons and phonons in the initial state can affect the QFI dependence on the total number of particles. In order to analyze the dependence of the QFI on the population, we introduce the population ratio
\begin{equation}\label{eq:14}
N_c=uN, \hspace{0.5cm} N_m=(1-u)N,
\end{equation}
where $N_c$ and $(N_m)$ are the number of photons (phonons) in the initial state of the QOMG. $N_c=|\alpha|^2$ and $N_m=|\beta|^2$ for $|\alpha\rangle|\beta\rangle$.

In. Fig.~\ref{fig:4}, we plot the QFI with respect to the total number of particles in the input state of the QOMG for different particle population when
we take the angular velocity $\Omega=2$kHz and the evolution time  $\tilde{\omega}_m t=2\pi$. The system's parameters of the QOMG are chosen as $g_{\text{NL}}=0.3g_0$, $\omega_c=10^{15}$Hz, $\omega_m=62.8$ kHz,  $L_0=10^{-4}$m , $D=10^{-3}$m, $m=10^{-7}$kg. The dot, dash, and  dash-dot line corresponds to the particle population  $u=0.1, 0.5$, and $0.9$, respectively. For the convenience of relevant comparisons, we also mark the standard quantum limit (SQL)  and the Heisenberg limit (HL) by the solid line and the triangle-solid line, respectively.
From Fig.~\ref{fig:4} we can observe that the sensitivity of the QOMG with Kerr interaction can achieve scaling beyond the HL, i.e., the super-Heisenberg limit. Specifically, for the values of the parameters given in Fig.~\ref{fig:4}, the sensitivity of the QOMG with Kerr interaction is several orders of magnitude higher than the HL. From Fig.~\ref{fig:4} we can also see that the particle population affects the sensitivity of the QOMG. The sensitivity increases with the photon population in the initial state. The improvement in sensitivity can exceed one order of magnitude when the photon population changes from $u=0.1$ to $u=0.9$.

	\section{\label{level3} Classical Fisher information in the quadrature measurement scheme}

In this section, we investigate  the sensitivity of the QOMG in a practical measurement scheme through  evaluating the classical Fisher information (CFI) with respect to the estimated parameter in a practical measurement scheme. We study the quadrature measurement of the cavity field in the QOMG without the driving field and show that the CFI in the quadrature measurement scheme can saturate  the QFI for the estimated parameter.

From Eq.~(\ref{eq:6}), the Hamiltonian of the QOMG without the driving field can be rewritten as
\begin{equation}\label{eq:15}
\begin{aligned}
H= & \hbar\left(\omega_c-\eta_0\right) n+\hbar \eta_0 n^2+\hbar \tilde{\omega}_m b^{\dagger} b \\
& +\hbar \tilde{\omega}_m\left[\left(\tilde{g}_0-\tilde{G}_{\mathrm{NL}}\right) n+\tilde{G}_{\mathrm{NL}} n^2-\tilde{\chi}\right]\left(b+b^{\dagger}\right),
\end{aligned}
\end{equation}
where $\tilde{g}_0=g_0 / \tilde{\omega}_m$, $\tilde{\chi}=\chi / \tilde{\omega}_m$, and $ \tilde{G}_{\mathrm{NL}}=G_{\mathrm{NL}} / \tilde{\omega}_m$.

We assume that the optical field in the cavity and the mechanical oscillator are initially in coherent states, $|\alpha\rangle$ and $|\beta\rangle$, respectively. Then, at time $t$, quantum state of the QOMG (see Appendix~\ref{appenB} for details) becomes
\begin{equation}
|\psi(t)\rangle=\sum_{n=0}^{\infty} e^{-\frac{|\alpha|^2}{2}} \frac{\alpha^n}{\sqrt{n!}} e^{i E_n^2 \tau(t)} e^{-i E_n \tilde{\beta}(t)}|n\rangle\left|\mu_n(t)\right\rangle,
\end{equation}
where $|n\rangle$ is a number state of the cavity field and $\left|\mu_n(t)\right\rangle$ is a coherent state of the mechanical oscillator with the amplitude $\mu_n(t)$. Parameters in Eq.~(\ref{eq:15}) are given by
\begin{equation}\label{eq:17}
\begin{aligned}
E_n=& \left(\tilde{g}_0-\tilde{G}_{\mathrm{NL}}\right) n+\tilde{G}_{\mathrm{NL}} n^2-\tilde{\chi}, \\
\mu_n(t)=&\beta e^{-i \tilde{\omega}_m t}-E_n\left(1-e^{-i \tilde{\omega}_{m } t}\right), \\
\tau(t)=&\tilde{\omega}_m t-\sin \tilde{\omega}_m t, \\
\tilde{\beta}(t)=&\beta_r \sin \tilde{\omega}_m t+\beta_i\left(1-\cos \tilde{\omega}_m t\right),\\
\end{aligned}
\end{equation}
where $\beta_r$ ($\beta_i$) is the real (imaginary) part of the initial-state parameter $\beta$.

From Eq.~(\ref{eq:15}), we can obtain the reduced density operator of the cavity field
\begin{equation}
\begin{aligned}
\rho_c(t)= & \sum_{n, n'} e^{-|\alpha|^2} \frac{\alpha^n}{\sqrt{n!}} \frac{\alpha^{*n'}}{\sqrt{n'!}}
e^{i\left(E_n^2-E_{n'}^{2}\right) \tau(t)} e^{-i\left(E_n-E_{n'}\right) \tilde{\beta}(t)} \\
& \times e^{-\frac{1}{2}\left|\mu_n(t)\right|^2-\frac{1}{2}\left|\mu_{n'}(t)\right|^2+\mu_n(t) \mu_{n'}^*(t)}|n\rangle\langle n'|.
\end{aligned}
\end{equation}

The quadrature operator of the cavity field is given by
\begin{equation}
X_\phi=\frac{1}{\sqrt{2}}\left(a e^{-i \phi}+a^{\dagger} e^{i \phi}\right),
\end{equation}
which has the eigenstate
\begin{equation}\label{eq:20}
|x\rangle_\phi=e^{-\frac{x^2}{2}}\left(\frac{1}{\pi}\right)^{\frac{1}{4}} \sum_m \frac{H_m(x)}{2^{\frac{m}{2}} \sqrt{m!}} e^{-i m \phi}|m\rangle,
\end{equation}
where $x$ is the eigenvalue of the quadrature operator  $X_\phi$, and  $H_m(x)$ is Hermite polynomial.

In the quadrature measurement scheme, the POVM is $V_x=|x\rangle_{\phi \phi}\langle x|$. Then, we can obtain  the  conditional probability distribution associated with the POVM  with the estimated parameter $\Omega$
\begin{equation}
P(x \mid \Omega)=  \operatorname{Tr}\left[\rho_c(t)|x\rangle_{\phi \phi}\langle x|\right].
\end{equation}

Substituting the reduced density operator of the cavity field given by Eq.~(\ref{eq:17}) into Eq.~(\ref{eq:20}), we can obtain the  conditional probability distribution
\begin{equation}
\begin{aligned}
P(x \mid \Omega)= & \frac{e^{-|\alpha|^2-x^2}}{\sqrt{\pi}} \sum_{n, n'} \frac{\alpha^n \alpha^{* n'}}{n!n'!2^{\frac{n+n'}{2}}} H_n(x) H_{n'}(x) \\
& \times e^{i\left(n-n^{\prime}\right) \phi} e^{i\left(E_n^2- E_{n'}^{2}\right) \tau} e^{-i\left(E_n-E_{n'}\right) \tilde{\beta}(t)} \\
& \times e^{-\frac{1}{2}\left|\mu_n(t)\right|^2-\frac{1}{2}\left|\mu_{n'}(t)\right|^2+\mu_n(t) \mu_{n'}^*(t)}.
\end{aligned}
\end{equation}

Making use of above  conditional probability distribution, we can calculate the CFI in the quadrature measurement scheme
\begin{equation}
F=\int \frac{\left[\partial_{\Omega} P(x \mid \Omega)\right]^2}{P(x \mid \Omega)} d x.
\end{equation}

\begin{figure}[t]
		\centerline{
			\includegraphics[width=0.4\textwidth]{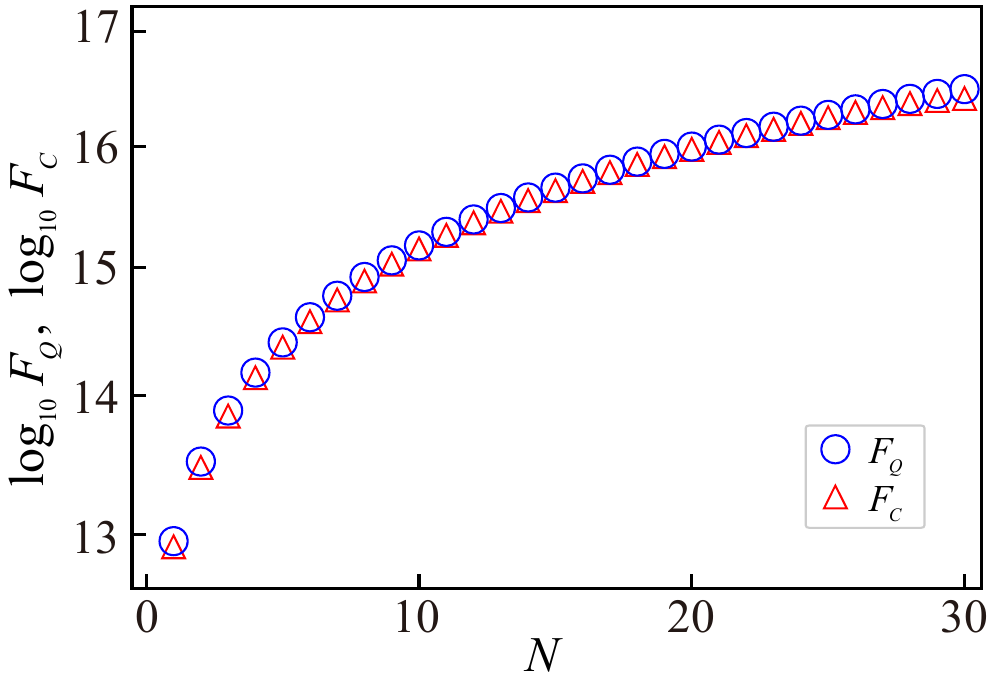}}
		\caption{The classical Fisher information on the quadrature measurement when $\Omega=2$kHz, $G_{\text{NL}}=0.3g_0$, and $u=0.9$.  The parameters of the QOMG system are chosen as $\omega_c=10^{15}$Hz, $\omega_m=62.8$ kHz,  $L_0=10^{-4}$m , $D=10^{-3}$m, $m=10^{-7}$kg. The initial state parameters are taken as $N_C=5$, $N_m=1$. The time is chosen as $\tilde{\omega}_m t=2\pi$.} \label{fig:5}
	\end{figure}

In Fig.~\ref{fig:5}, we plot the CFI in the quadrature measurement scheme with the total number of particles by the triangle line when $\Omega=2$kHz, $G_{\text{NL}}=0.3g_0$, and $u=0.9$.  The system's parameters of the QOMG are chosen as $\omega_c=10^{15}$Hz, $\omega_m=62.8$ kHz,  $L_0=10^{-4}$m , $D=10^{-3}$m, $m=10^{-7}$kg. The initial state parameters are taken as $N_C=5$, $N_m=1$. The time is chosen as $\tilde{\omega}_m t=2\pi$.
In order to compare the CFI with  the QFI,  in Fig.~\ref{fig:5}, we also plot QFI with the total number of particles by the circle line under the same parameter condition. From Fig.~\ref{fig:5}, we can observe that the triangle curve and the circle line almost overlap. This means that the CFI completely approaches the QFI.  Hence, we can conclude that the sensitivity of the QOMG with Kerr interaction in the quadrature measurement scheme can saturate  the quantum Cram\'{e}r-Rao bound.

\section{\label{level3} Influence of driving and dissipation}

In this section, we numerically investigate influence of driving and dissipation of the cavity field on  the QOMG sensitivity using the quantum toolbox QuTIP  for numerical solutions \cite{Johansson}.
We  shall indicate  that  the  QOMG sensitivity can be manipulated by changing the amplitude or/and phase of the driving field  while  dissipation decreases the sensitivity.

\begin{figure}[t]
	\centerline{
		\includegraphics[width=0.4\textwidth]{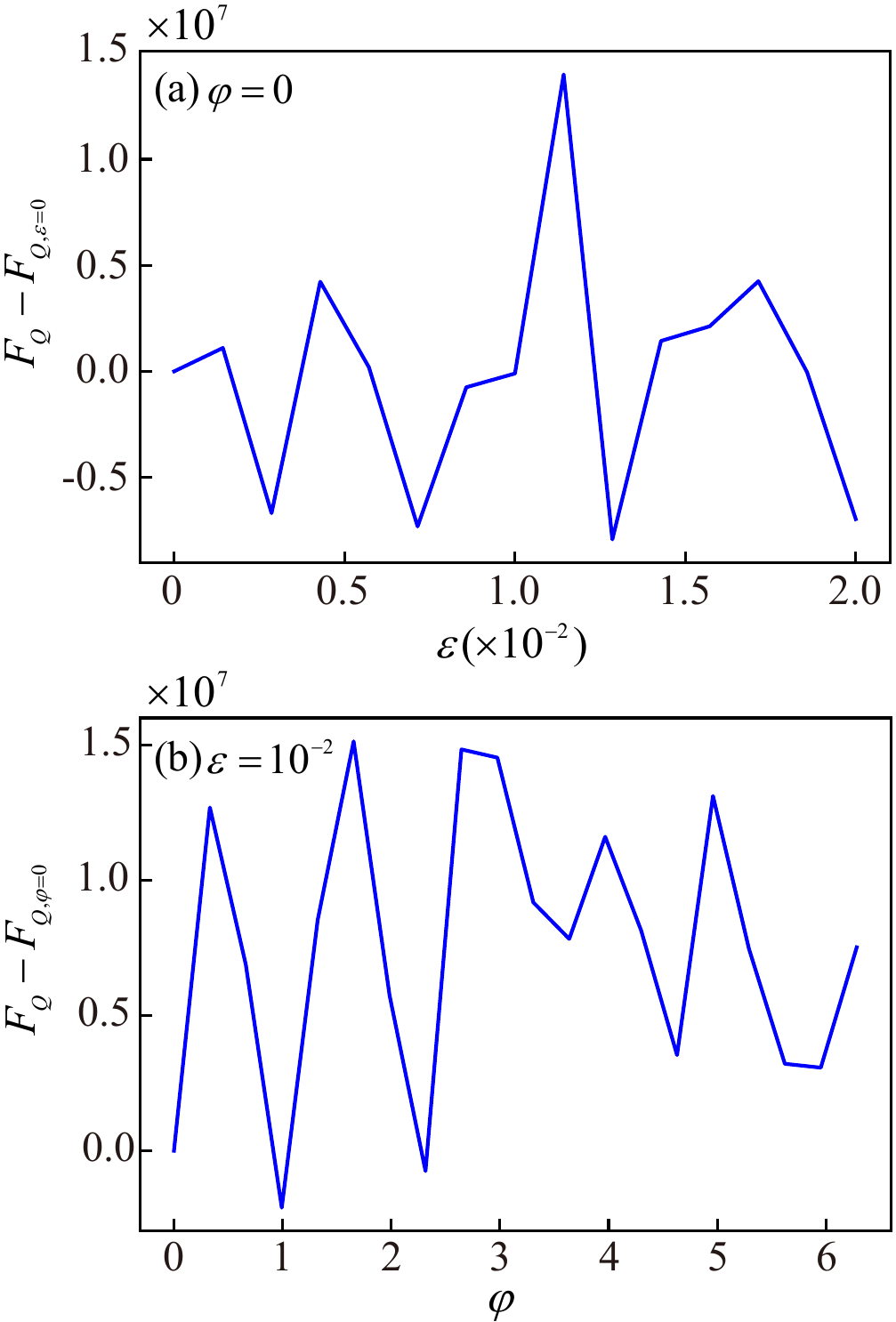}}
	\caption{The QFI as a function of parameters of the driving field. (a) The amplitude driving with the fixed phase  $\varphi=0$. (b) The phase driving with the fixed driving amplitude  $\varepsilon=1\times10^{-2}$.  The parameters are $\Omega=2\times10^{-12}$ , $G_{\text{NL}}=0.3g_0$.  The  dimensionless parameters of the QOMG system  are chosen as $\omega_c=1$, $\omega_m=62.8\times10^{-12}$, and  $L_0=10^{-4}$m , $D=10^{-3}$m, $m=10^{-7}$kg. The initial state parameters are taken as $N_C=5$, $N_m=1$. The evolution dimensionless time takes $T=2\pi$.} \label{fig6}
\end{figure}

We now examine the effect of the driving on the QOMG sensitivity within our framework. The corresponding Hamiltonian is expressed in Eq.~(\ref{eq:6}). Based on the Hamiltonian, the quantum state of the system at any time can be numerically obtained. Using the definition of the QFI for pure states Eq.~(\ref{eq:8}), we analyze how the QFI varies with the driving strength. To simplify the analysis, we introduce dimensionless units by setting $\hbar=1$ and rescaling all frequencies with respect to the cavity frequency $\omega_{c}=10^{15}$Hz. This yields $\omega_c = 1$ in dimensionless units, and, for example, the mechanical frequency becomes $\omega_m = 62.8 \times 10^{-12}$ and the angular velocity of QOMG becomes $\Omega=2\times10^{-12}$.  Time is expressed in units of $\omega_{c}^{-1}$, and the dimensionless evolution time is denoted as $T = \omega_c t$, where $\omega_c$ and $t$ refer to the dimensional (physical) cavity frequency and time, respectively. The other physical parameters are chosen as $L_{0}=10^{-4}\text{m}$, $D=10^{-3}\text{m}$, and $m=10^{-7}\text{kg}$. The system is initialized in the product state $\left|\alpha\right\rangle\left|\beta\right\rangle$ with $\alpha=\sqrt{5}$ and $\beta=1$, i.e., $N_{c}=5$ and $N_{m}=1$. We compute the QFI at the dimensionless time point $T=2\pi$ for varying values of the driving strength $\varepsilon=|\varepsilon|e^{i\varphi}$  to observe its influence on the sensitivity of the system. In Fig.~\ref{fig6}(a), we investigate the effect of the driving amplitude on the QFI, with the driving phase fixed at $\varphi=0$. The results show that the QFI exhibits oscillatory variations with respect to the driving amplitude. Fig.~\ref{fig6}(b), we explore the influence of the driving phase on the QFI, with the driving amplitude set to $\varepsilon=1\times10^{-2}$. We can find the QFI exhibits oscillatory variations with respect to the driving phase. However the overall influence of the driving on the QFI remains relatively small. The QFI again shows an oscillatory dependence, this time with respect to the driving phase. However, in both cases, the overall impact of the external driving on the QFI remains relatively limited.

In realistic scenarios, environmental effects are inevitable, and noise can also influence the sensitivity. Here we consider dissipation of the cavity on the QOMG sensitivity. For open quantum systems, a quantum state is typically described by a density operator, denoted as $\rho$, whose dynamical evolution is governed by the Lindblad master equation. The general form of the Lindblad master equation is given by
\begin{equation}
	\frac{d\rho}{dt}=-\frac{i}{\hbar}\left[H,\rho\right]+\sum_{k}\left(L_{k}\rho L_{k}^{\dagger}-\frac{1}{2}\left\{L_{k}^{\dagger}L_{k},\rho\right\}\right)
\end{equation}
where $H$ is the system Hamiltonian and $L_{k}$ are the Lindblad operators describing the dissipative channels. The index $k$ enumerates the distinct types of coupling between the system and its environment. In our model, we focus solely on the dissipation of the optical mode, since the mechanical damping rate is typically much smaller than the optical loss rate and can be safely neglected \cite{G.-L. Li}. The relevant dissipation is described by a single Lindblad operator $L=\sqrt{\kappa}a$, where $\kappa$ is the cavity decay rate and $a$ is the annihilation operator of the cavity field. Based on this open-system framework, we can numerically simulate the time evolution of the system under cavity decay.  Since the evolved state of an open system is generally a mixed state, the QFI dynamics can be evaluated using the formalism for mixed states \cite{J. Liu}
\begin{equation}
	\begin{aligned}
	F_{Q}=&\sum_{i=1}^{S}\frac{(\partial_{\Omega}P_{i})^{2}}{P_{i}}+\sum_{i=1}^{S}4P_{i}\left\langle \partial_{\Omega}\Psi_{i}|\partial_{\Omega}\Psi_{i}\right\rangle \\
	&-\sum_{i,j=1}^{S}\frac{8P_{i}P_{j}}{P_{i}+P_{j}}\left|\left\langle\Psi_{i}|\partial_{\Omega}\Psi_{j}\right\rangle\right|^{2}
	\end{aligned}
\end{equation}
where $P_{i}$ and $\left|\Psi_{i}\right\rangle$ denote the $i$-th eigenvalue and the corresponding eigenstate of the density operator $\rho(t)$ at time $t$, respectively.

\begin{figure}[t]
	\centerline{
		\includegraphics[width=0.4\textwidth]{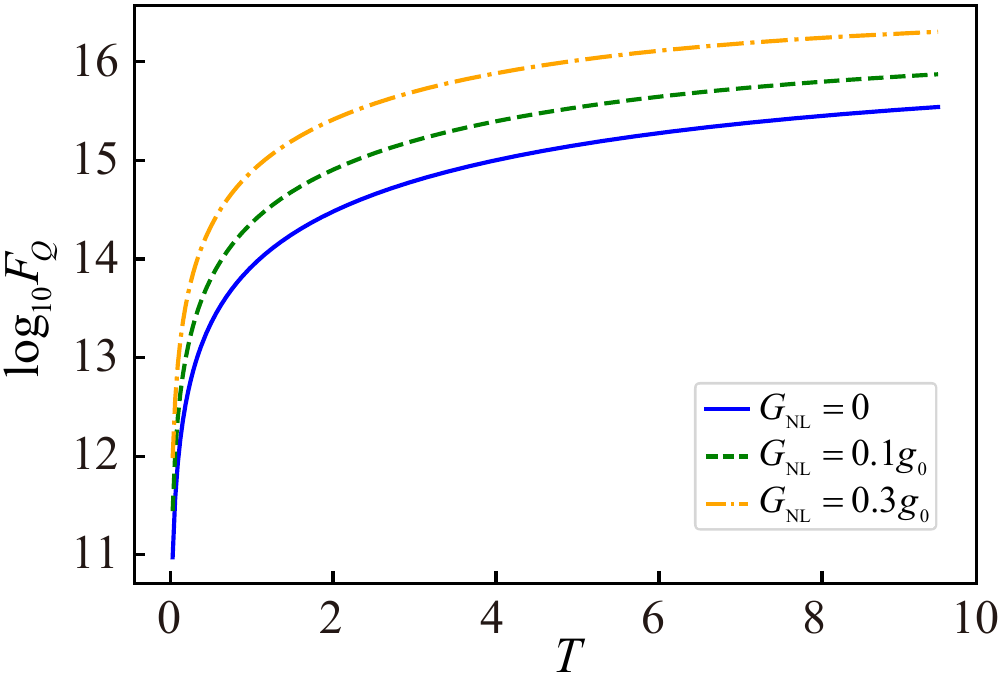}}
	\caption{Dynamic evolution of the QFI for different values of the nonlinear coupling strength when $\Omega=2\times10^{-12}$, and  $\kappa=0.1$.  The dimensionless parameters of the QOMG  system are chosen as $\omega_c=1$, $\omega_m=62.8\times10^{-12}$, and  $L_0=10^{-4}$m , $D=10^{-3}$m, $m=10^{-7}$kg. The initial state parameters are taken as $N_C=5$, $N_m=1$.} \label{fig7}
\end{figure}

Consequently, we numerically obtain the dynamical evolution of the QFI with respect to the rotational angular velocity in the presence of dissipation. In Fig.~\ref{fig7}, we present the dynamical evolution of the QFI for different values of the Kerr-type nonlinear coupling strength, under a fixed dissipation rate of $\kappa=0.1$. All other system parameters are kept constant. The results demonstrate that even in the presence of optical dissipation, the sensitivity of the QOMG scheme can be further enhanced as the nonlinear interaction strength increases.
\begin{figure}[t]
	\centerline{
		\includegraphics[width=0.4\textwidth]{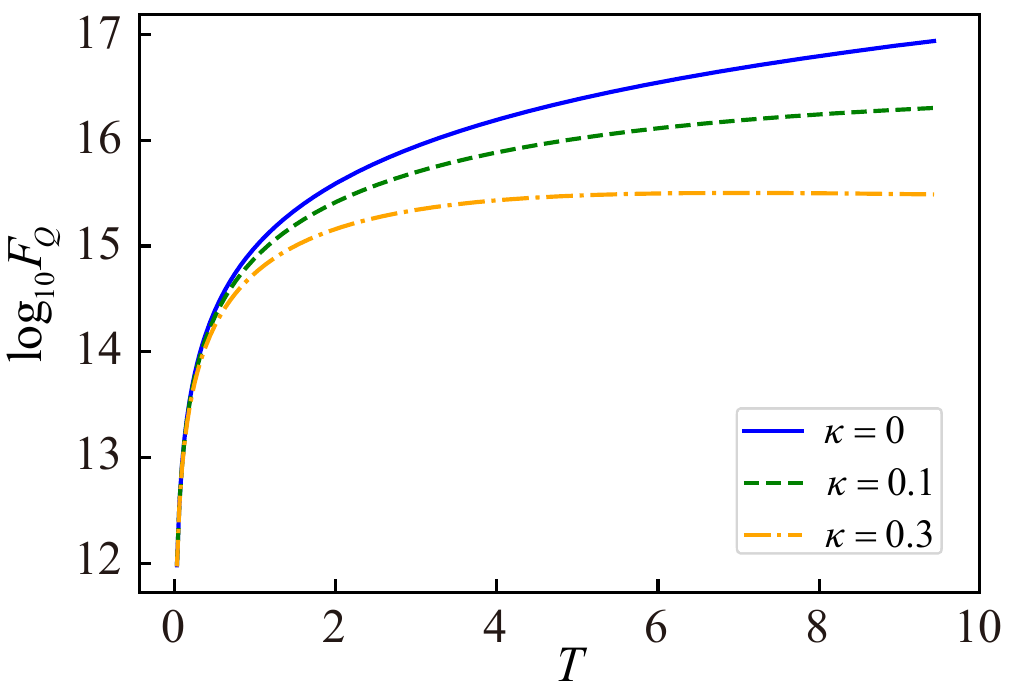}}
	\caption{Dynamic evolution of the QFI for different dissipation rates of the cavity  when $\Omega=2\times10^{-12}$, and $G_{\text{NL}}=0.3g_0$.  The  dimensionless parameters of the QOMG  system are chosen as $\omega_c=1$, $\omega_m=62.8\times10^{-12}$ , and  $L_0=10^{-4}$m , $D=10^{-3}$m, $m=10^{-7}$kg. The initial state parameters are taken as $N_C=5$, $N_m=1$.} \label{fig8}
\end{figure}

Furthermore, we investigate the effect of dissipation on the sensitivity by fixing the nonlinear interaction strength at $G_{\text{NL}}=0.3g_{0}$ and varying the dissipation rates as $\kappa=0, \kappa=0.1, \kappa=0.3$. The analysis is performed under the same dimensionless parameters: $\hbar=1$, $\omega_{c}=1$, $\omega_{m}=62.8\times10^{-12}$, $\Omega=2\times10^{-12}$, and  $L_{0}=10^{-4}\text{m}$, $D=10^{-3}\text{m}$, $m=10^{-7}\text{kg}$. The corresponding QFI dynamics are shown in Fig.~\ref{fig8}. It is observed that dissipation has a detrimental effect on the sensitivity, with the QFI decreasing as the dissipation rate increases. Nevertheless, even in the presence of dissipation, the QFI remains at a relatively high magnitude, indicating that the scheme retains robustness against moderate losses.

\begin{figure}[t]
	\centerline{
		\includegraphics[width=0.4\textwidth]{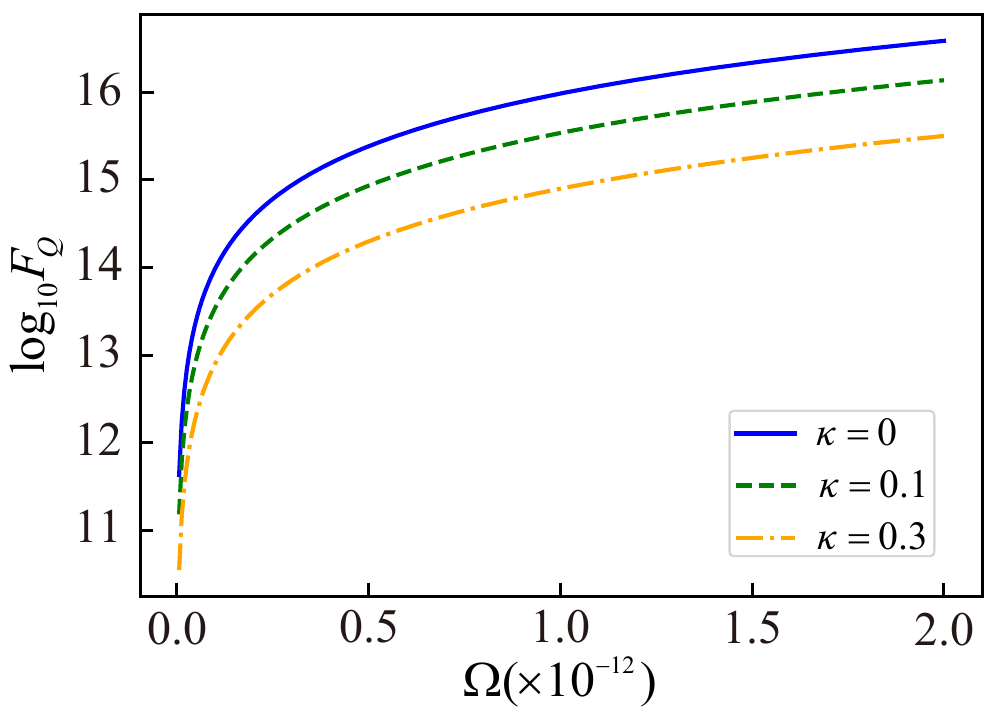}}
	\caption{The QFI with respect to the angular velocity of the QOMG. The  dimensionless parameters of the QOMG  system are chosen as $\omega_c=1$, $\omega_m=62.8\times10^{-12}$, and  $L_0=10^{-4}$m , $D=10^{-3}$m, $m=10^{-7}$kg. The initial state parameters are taken as $N_C=5$, $N_m=1$. The evolution dimensionless time obeys $T=2\pi$.} \label{fig9}
\end{figure}

Similarly, we also examine the operational regime of the QOMG scheme in the presence of dissipation. As shown in Fig.~\ref{fig9}, we plot the QFI as a function of the QOMG angular velocity under different dissipation rates, with the dimensionless evolution time fixed at $T=2\pi$ and other parameters remain unchanged. We find that, in the presence of dissipation, the sensitivity of the QOMG is consistently reduced across the entire range of angular velocities. Moreover, as the dissipation rate increases, the degradation of sensitivity becomes more pronounced.

\begin{figure}[t]
	\centerline{
		\includegraphics[width=0.4\textwidth]{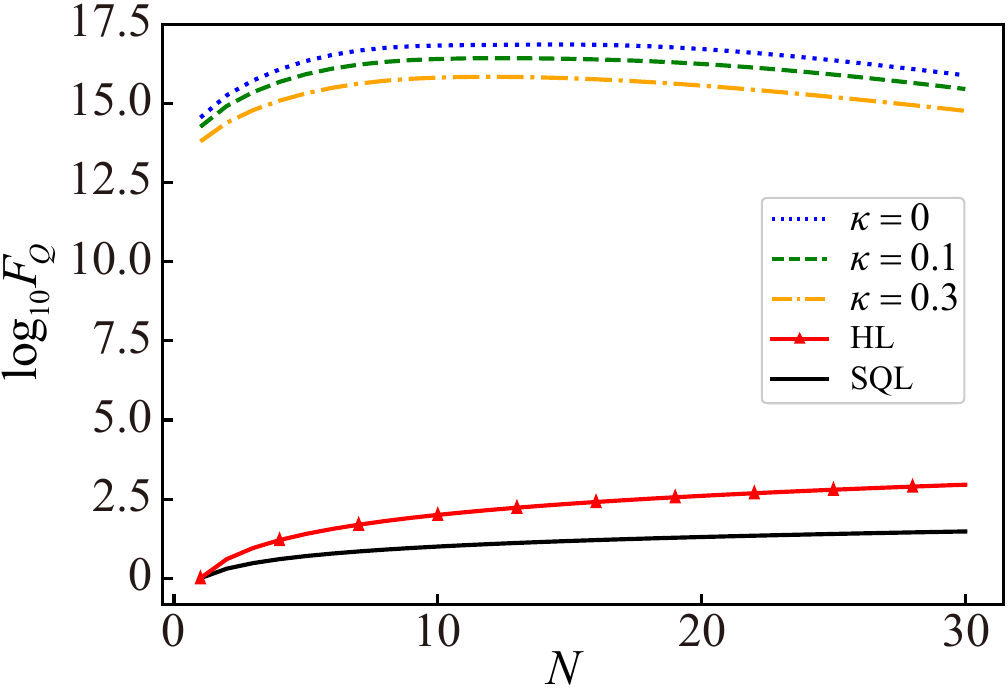}}
	\caption{The QFI with respect to total number of particles in the input state of the QOMG, when $\Omega=2\times10^{-12}$, $g_{\text{NL}}=0.3g_0$ and $u=0.9$. The   dimensionless parameters of the QOMG  system  are chosen as $\omega_c=1$, $\omega_m=62.8\times10^{-12}$, and  $L_0=10^{-4}$m , $D=10^{-3}$m, $m=10^{-7}$kg. The initial state parameters are taken as $N_C=5$, $N_m=1$. The evolution dimensionless time $T=2\pi$} \label{fig10}
\end{figure}

Finally, we examine how the QFI scales with the input resource number in the presence of dissipation.  As shown in Fig.~\ref{fig10}, we plot the QFI versus resource number under three different dissipation rates, $\kappa=0, 0.1, 0.3$,  represented by dashed, dotted, and dash-dotted lines, respectively. The angular velocity is fixed at $\Omega=2\times10^{-12}$, the evolution dimensionless time is set to $T=2\pi$, the nonlinear interaction strength is $G_{\text{NL}}=0.3g_{0}$, and the population ratio is $u=0.9$. And to investigate whether the sensitivity under dissipation can still surpass the SQL and the HL, we also plot the SQL and HL as references, indicated by the solid line and triangular-solid line, respectively. It can be seen that although dissipation reduces the sensitivity, the QFI still remains several orders of magnitude above the Heisenberg limit. This indicates that, to some extent, our scheme exhibits robustness against optical losses. This result highlights the potential practicality of the angular velocity sensing scheme using QOMG with kerr interaction under realistic experimental conditions where optical losses are unavoidable.

	\section{\label{level5}Concluding remarks}
	
In this work, we have proposed  a theoretical scheme to enhance the sensitivity of the QOMG  by optical Kerr interaction. We have evaluated the
metrological potential of the QOMG scheme by analyzing quantum Fisher information.
It has been indicated that  the Kerr interaction can significantly enhances the sensitivity of the QOMG.
We have observed the super-Hesenberg scaling of parameter estimation precision in the QOMG.  Specifically, for the certain values of the parameters, the sensitivity of the QOMG with Kerr interaction is several orders of magnitude higher than the Hesenberg limit. We have also found that the photon and phonon population in the initial state can affect the sensitivity of the QOMG. Under the condition with the same total number of particles, the sensitivity increases with the photon population in the initial state.
Furthermore, we have explored the performance of QOMG in a practical quantum measurement scheme, the quadrature measurement. It han been shown that the sensitivity in the quadrature measurement scheme can saturate the quantum Cram\'{e}r-Rao bound. This implies that  the quadrature measurement is an optimal measurement for the QOMG.
Finally, we have studied the effect of the driving and dissipation of the optical cavity on  the QFI of the QOMG. We have found that  the sensitivity can be manipulated by changing the driving field while  dissipation decreases the sensitivity. The work shows that the  photon Kerr interaction can  significantly improve  sensitivity of QOMG, and demonstrates a valuable quantum resource for the QOMG. These results could have a wide-ranging impact on developing  high-performance QOMG in the future.

\begin{acknowledgements}
 L.-M. K. is supported by the Natural Science Foundation of China (NSFC) (Grant Nos. 12247105, 12175060, 12421005), the Innovation Program for Quantum Science and Technology (Grant No. 2024ZD0301000), Hunan provincial sci-tech program (Grant No. 2023ZJ1010) and  XJ-Lab key project (Grant No. 23XJ02001).  Y.-F.J. is supported by the NSFC (Grant No. 12405029) and the Natural Science Foundation of Henan Province (Grant No. 252300421221). W.-J.L.  is supported by the NSFC (Grant No. 12205092).
 Q.-S.T.  is supported by the NSFC (Grant No. 12275077).

	\end{acknowledgements}

\appendix
\section{\label{appenA}Derivation of Equations (12) and (13)}

In the Appendix A, we present derivation of Equations~(\ref{eq:12}) and~(\ref{eq:13}) in the second section.
The generator of parameter translation can be expanded in terms of the Hamiltonian of the system as
\begin{equation}\label{A1}
\mathcal{H}=i \sum_{k=0}^{\infty} \frac{(i t / \hbar)^{k+1}}{(k+1)!} H^{\times k}\left(\partial_{\Omega} H\right),
\end{equation}

Making use of Eq.~(\ref{eq:6}), in the absence of the driving ($\epsilon=0$) we can obtain
\begin{equation}\label{A2}
\begin{aligned}
\partial_{\Omega} H= & \frac{\hbar \Omega}{\tilde{\omega}_m} b^{\dagger} b-\frac{\hbar \Omega}{2 \tilde{\omega}_m^2}\left[G_{\mathrm{NL}} n^2+\left(g_0-G_{\mathrm{NL}}\right) n\right. \\
& \left.+\left(\frac{4 \tilde{\omega}_m^2}{\Omega^2}-1\right) \chi\right]\left(b+b^{\dagger}\right),
\end{aligned}
\end{equation}
where $n=a^{\dagger}a$ is the number operator of photons.

It is easy to calculate first-order commutator between $H$ and $\partial_{\Omega}H$ with the following expression
\begin{equation}\label{A3}
\begin{aligned}
H^{\times 1}\left(\partial_{\Omega} H\right)= & \frac{\hbar^2 \Omega}{2 \tilde{\omega}_m}\left[3 G_{\mathrm{NL}} n^2+3\left(g_0-G_{\mathrm{NL}}\right) n\right. \\
& \left.+\left(\frac{4 \tilde{\omega}_m^2}{\Omega^2}-3\right) \chi\right]\left(b-b^{\dagger}\right),
\end{aligned}
\end{equation}
and the second-order  commutator between $H$ and $\partial_{\Omega}H$ is given by
\begin{equation}\label{A4}
\begin{aligned}
H^{\times 2}\left(\partial_{\Omega} H\right)= & -\frac{\hbar^3 \Omega}{2}\left[3 G_{\mathrm{NL}} n^2+3\left(g_0-G_{\mathrm{NL}}\right) n\right. \\
& \left.+\left(\frac{4 \tilde{\omega}_m^2}{\Omega^2}-3\right) \chi\right] \times\left\{\left(b+b^{\dagger}\right)\right. \\
& \left.+\frac{2}{\tilde{\omega}_m}\left[g_0 n-\chi+G_{\mathrm{NL}}\left(n^2-n\right)\right]\right\}.
\end{aligned}
\end{equation}
In general, a high-order  commutator  can be recursively derived using a low-order commutator through the following recurrence relation
\begin{equation}\label{A5}
\begin{aligned}
& H^{\times(2 k+1)}\left(\partial_{\Omega} H\right)=\left(\hbar \tilde{\omega}_m\right)^{2 k} H^{\times 1}\left(\partial_{\Omega} H\right), \\
& H^{\times(2 k+2)}\left(\partial_{\Omega} H\right)=\left(\hbar \tilde{\omega}_m\right)^{2 k} H^{\times 2}\left(\partial_{\Omega} H\right),
\end{aligned}
\end{equation}
where $k$ takes a non negative integer.

Making usee of Eq.~(\ref{A5}), we can rewrite Eq.~(\ref{A1}) as
\begin{equation}\label{A6}
\begin{aligned}
\mathcal{H}= & -\frac{t}{\hbar} \partial_{\Omega} H+i \sum_{k=0}^{\infty} \frac{(i \tilde{\omega}_m t)^{2 k+2}}{(2 k+2)!} \frac{H^{\times 1}\left(\partial_{\Omega} H\right)}{\hbar^2 \tilde{\omega}_m^2} \\
& +i \sum_{k=0}^{\infty} \frac{(i \tilde{\omega}_m t)^{2 k+3}}{(2 k+3)!} \frac{H^{\times 2}\left(\partial_{\Omega} H\right)}{\hbar^3 \tilde{\omega}_m^3},
\end{aligned}
\end{equation}
which can be simplified as
\begin{equation}\label{A7}
\begin{aligned}
\mathcal{H}= & -\frac{t}{\hbar} \partial_{\Omega} H+i\left(\cos \tilde{\omega}_m t -1\right) \frac{H^{\times 1}\left(\partial_{\Omega} H\right)}{\hbar^2 \tilde{\omega}_m^2} \\
& -\left(\sin \tilde{\omega}_m t - \tilde{\omega}_m t\right) \frac{H^{\times 2}\left(\partial_{\Omega} H\right)}{\hbar^3 \tilde{\omega}_m^3}.
\end{aligned}
\end{equation}

Substituting Eq.~(\ref{A3}) and  Eq.~(\ref{A4}) into Eq.~(\ref{A7}), we can express Eq.~(\ref{A7}) as
\begin{equation}\label{A8}
\mathcal{H}=-\frac{\Omega}{2 \tilde{\omega}_m^3}\left(\mathcal{H}_1+\mathcal{H}_2+\mathcal{H}_3+\mathcal{H}_4+\mathcal{H}_5\right),
\end{equation}
which is Eq.~(\ref{eq:12}) in the main text. Here $\mathcal{H}_i$ is defined by
\begin{equation}\label{A9}
\begin{aligned}
\mathcal{H}_1 & =R_1 n^4, \hspace{0.5cm}
\mathcal{H}_2  =R_2  n^3 ,\\
\mathcal{H}_3 & =(R_3  b+R_3^*  b^{\dagger}+R_4)  n^2 , \\
\mathcal{H}_4 & =(R_5  b+R_5^*  b^{\dagger}+R_6) n, \\
\mathcal{H}_5 & =R_7  b^{\dagger} b+R_8  b+R_8^*  b^{\dagger},
\end{aligned}
\end{equation}
where $n=a^{\dagger}a$ is the number operator of photons and  $R_i$ coefficients are given by
\begin{equation}
\begin{aligned}
& R_1 =6 \frac{G_{\mathrm{NL}}^2}{\tilde{\omega}_m} C_1 , \\
& R_2 =12 \frac{G_{\mathrm{NL}}}{\tilde{\omega}_m}\left(g_0-G_{\mathrm{NL}}\right) C_1 , \\
& R_3 =3 G_{\mathrm{NL}} C_2 -G_{\mathrm{NL}} \tilde{\omega}_m t, \\
& R_4 =\frac{2 C_1 }{\tilde{\omega}_m}\left[3\left(g_0-G_{\mathrm{NL}}\right)^2+\left(4 \frac{\tilde{\omega}_m^2}{\Omega^2}-6\right) \chi G_{\mathrm{NL}}\right], \\
& R_5 =\left(g_0-G_{\mathrm{NL}}\right)\left[3 C_2 -\tilde{\omega}_m t\right], \\
& R_6 =\left(g_0-G_{\mathrm{NL}}\right)\left(\frac{8 \tilde{\omega}_m}{\Omega^2}-\frac{12}{\tilde{\omega}_m}\right) \chi C_1 , \\
& R_7 =2 \tilde{\omega}_m t, \\
& R_8 =\left[4 \frac{\tilde{\omega}_m^2}{\Omega^2}\left(C_2 -\tilde{\omega}_m t\right)-3 C_2 +\tilde{\omega}_m t\right] \chi,
\end{aligned}
\end{equation}
where $C_i$ coefficients are defined as
\begin{equation}
\begin{aligned}
& C_1 =\tilde{\omega}_m t -\sin \tilde{\omega}_m t, \\
& C_2 =(\tilde{\omega}_m t -\sin \tilde{\omega}_m t)+i(1-\cos \tilde{\omega}_m t).
\end{aligned}
\end{equation}

Substituting Eq.~(\ref{A8}) into the expression of the QFI given by Eq.~(\ref{eq:8}), we can get
\begin{equation}
\mathcal{F}=\frac{\Omega^2}{\tilde{\omega}_m^6}\left[\sum_{i=1}^5 \operatorname{Var}\left(\mathcal{H}_i\right)+\sum_{j, k=1, j \neq k}^5 \operatorname{Cov}\left(\mathcal{H}_j, \mathcal{H}_k\right)\right],
\end{equation}
This is Eq.~(\ref{eq:13}) in the main text. Here the variance of $\mathcal{H}_i$ and the  covariance matrix are defined by
\begin{equation}
\begin{aligned}
& \operatorname{Var}(\mathcal{H}_i)=\left\langle \mathcal{H}_i^2\right\rangle-\langle \mathcal{H}_i\rangle^2, \\
& \operatorname{Cov}\left(\mathcal{H}_i, \mathcal{H}_j\right)=\frac{1}{2}\left\langle\left\{\mathcal{H}_i, \mathcal{H}_j\right\}\right\rangle-\left\langle \mathcal{H}_i\right\rangle\langle \mathcal{H}_j\rangle.
\end{aligned}
\end{equation}

Assume that the optical field in the cavity and the mechanical oscillator are initially in  the coherent state $|\alpha, \beta\rangle$ with the the number of  photons $N_c=|\alpha|^2$ and  the the number of  phonons $N_m=|\beta|^2$, respectively. Making use of Eq.~(\ref{A9}), we can calculate the variances and the covariance matrix of  $\mathcal{H}_i$. The variances  of  $\mathcal{H}_i$ are given by
\begin{equation}
\begin{aligned}
\operatorname{Var}\left(\mathcal{H}_1\right)= & R_1^2 \left(16 N_c^7+216 N_c^6+964 N_c^5+1640 N_c^4\right. \\
& \left.+952 N_c^3+126 N_c^2+N_c\right), \\
\operatorname{Var}\left(\mathcal{H}_2\right)= & R_2^2 \left(9 N_c^5+54 N_c^4+84 N_c^3+30 N_c^2+N_c\right), \\
\operatorname{Var}\left(\mathcal{H}_3\right)= & \left\{\left(R_3 +R_3^* \right) N_m^{\frac{1}{2}}+R_4 \right\}^2\left(N_c+6 N_c^2+4 N_c^3\right) \\
& +\left|R_3 \right|^2\left(N_c+7 N_c^2+6 N_c^3+N_c^4\right), \\
\operatorname{Var}\left(\mathcal{H}_4\right)= & \left\{\left[\left(R_5 +R_5^* \right) N_m^{\frac{1}{2}}+R_6 \right]^2+\left|R_5 \right|^2\right\} N_c +\left|R_5 \right|^2 N_c^2, \\
\operatorname{Var}\left(\mathcal{H}_5\right)= & R_7^2  N_m+R_7 \left(R_8 +R_8^* \right) N_m^{\frac{1}{2}}+\left|R_8 \right|^2,
\end{aligned}
\end{equation}
And the the elements of the covariance matrix are given by
\begin{equation}
\begin{aligned}
&\begin{aligned}
\operatorname{Cov}\left(\mathcal{H}_1, \mathcal{H}_2\right)= & R_1  R_2  \times\left(12 N_c^6+114 N_c^5+322 N_c^4+291 N_c^3\right. \\
& \left.+62 N_c^2+N_c\right), \\
\operatorname{Cov}\left(\mathcal{H}_1, \mathcal{H}_3\right)= & R_1 \left[\left(R_3 +R_3^* \right) N_m^{\frac{1}{2}}+R_4 \right] \\
& \times\left(8 N_c^5+52 N_c^4+82 N_c^3+30 N_c^2+N_c\right), \\
\operatorname{Cov}\left(\mathcal{H}_1, \mathcal{H}_4\right)= & R_1 \left[\left(R_5 +R_5^* \right) N_m^{\frac{1}{2}}+R_6 \right] \\
& \times\left(4 N_c^4+18 N_c^3+14 N_c^2+N_c\right), \\
\operatorname{Cov}\left(\mathcal{H}_1, \mathcal{H}_5\right)= & 0, \\
\operatorname{Cov}\left(\mathcal{H}_2, \mathcal{H}_3\right)= & R_2 \left[\left(R_3 +R_3^* \right) N_m^{\frac{1}{2}}+R_4 \right] \\
& \times\left(6 N_c^4+21 N_c^3+14 N_c^2+N_c\right),
\end{aligned}\\
&\begin{aligned}
\operatorname{Cov}\left(\mathcal{H}_2, \mathcal{H}_4\right)= & R_2 \left[\left(R_5 +R_5^* \right) N_m^{\frac{1}{2}}+R_6 \right]\left(3 N_c^3+6 N_c^2+N_c\right), \\
\operatorname{Cov}\left(\mathcal{H}_2, \mathcal{H}_5\right)= & 0, \\
\operatorname{Cov}\left(\mathcal{H}_3, \mathcal{H}_4\right)= & \left[\left(R_3 +R_3^* \right)\left(R_5 +R_5^* \right) N_m +R_6 \left(R_3 +R_3^* \right) N_m^{\frac{1}{2}} \right.\\
& \left.+R_4  \left(R_5 +R_5^* \right)   N_m^{\frac{1}{2}}+R_4  R_6 \right] \left(2 N_c^2+N_c\right) \\
& +\frac{1}{2}\left(R_3  R_5^* +R_3^*  R_5 \right)\left(N_c^3+3 N_c^2+N_c\right),
\end{aligned}\\
&\begin{aligned}
\operatorname{Cov}\left(\mathcal{H}_3, \mathcal{H}_5\right)= & \frac{1}{2}\left[\left(R_3 +R_3^* \right) R_7  N_m^{\frac{1}{2}}+R_3  R_8^* +R_3^*  R_8 \right]\left(N_c^2+N_c\right),
\end{aligned}\\
&\begin{aligned}
\operatorname{Cov}\left(\mathcal{H}_4, \mathcal{H}_5\right)= & \frac{1}{2} N_c \left[\left(R_5 +R_5^* \right) R_7  N_m^{\frac{1}{2}}+R_5  R_8^* +R_5^*  R_8 \right].
\end{aligned}
\end{aligned}
\end{equation}

\section{\label{appenB}Derivation of Equation (15)}

In the Appendix B, we present derivation of Equation~(\ref{eq:15}) in the second section.  The Hamiltonian of the QOMG without the driving field is given by Eq.~(\ref{eq:14}) in the main text
\begin{equation}
\begin{aligned}
H= & \hbar\left(\omega_c-\eta_0\right) n+\hbar \eta_0 n^2+\hbar \tilde{\omega}_m b^{\dagger} b \\
& +\hbar \tilde{\omega}_m\left[\left(\tilde{g}_0-\tilde{G}_{\mathrm{NL}}\right) n+\tilde{G}_{\mathrm{NL}} n^2-\tilde{\chi}\right]\left(b+b^{\dagger}\right).
\end{aligned}
\end{equation}

In the interaction picture associated with the unitary transformation $U_0=e^{-i t\left[\left(\omega_{c}-\eta_0\right) n+\eta_0 n^2\right]}$, the transformed Hamiltonian becomes
\begin{equation}
\begin{aligned}
H'& =i \hbar \frac{d U_0^{\dagger}}{d t} U_0+U_0^{\dagger} H U_0 \\
& =\hbar \tilde{\omega}_m b^{\dagger} b+\hbar \tilde{\omega}_m\left[\left(\tilde{g}_0-\tilde{G}_{\mathrm{NL}}\right) n+\tilde{G}_{\mathrm{NL}} n^2-\tilde{\chi}\right]\left(b+b^{\dagger}\right).
\end{aligned}
\end{equation}

The unitary evolution operator associated with the Hamiltonian $H'$ can be written as
\begin{equation}
U(t)= e^{-i \tilde{\omega}_m t\left[b^{\dagger} b+E_n\left(b+b^{\dagger}\right)\right]},
\end{equation}
where we have introduced the following expression
\begin{equation}
E_n=\left(\tilde{g}_0-\tilde{G}_{\mathrm{NL}}\right) n+\tilde{G}_{\mathrm{NL}} n^2-\tilde{\chi}.
\end{equation}

Making the displacement transformation $D=e^{E_n\left(b^{\dagger}-b\right)}$ on the unitary operator $U(t)$, we obtain
\begin{equation}
D U(t) D^{\dagger}=e^{-i \tilde{\omega}_m t\left(b^{\dagger} b-E_n^2\right)}.
\end{equation}
which leads to
\begin{equation}\label{B6}
\begin{aligned}
U(t) & =e^{i \tilde{\omega}_m t E_n^2} e^{-E_n\left(b^{\dagger}-b\right)} e^{-i \tilde{\omega}_m t b^{\dagger} b} e^{E_n\left(b^{\dagger}-b\right)}\\
     & =e^{i \tilde{\omega}_m t E_n^2} e^{-E_n\left(b^{\dagger}-b\right)} e^{E_n\left(b^{\dagger} e^{-i \tilde{\omega}_m t}-b e^{i \tilde{\omega}_m t}\right)} e^{-i \tilde{\omega}_m t b^{\dagger} b}.
\end{aligned}
\end{equation}

By the use of the Baker-Hausdorff formula, we can re-express Eq.~(\ref{B6}) as
\begin{equation}
U(t) =e^{i E_n^2 \tau} e^{-E_n\left[\lambda(t) b^{\dagger}-\lambda^*(t) b\right]} e^{-i \tilde{\omega}_m t b^{\dagger} b},
\end{equation}
where $\tau(t)=\tilde{\omega}_m t-\sin \tilde{\omega}_m t$ and $\lambda(t) =1-e^{-i \tilde{\omega}_{m} t}$.

Assume that the optical field   and the mechanical oscillator are initially in the coherent state, $|\alpha\rangle|\beta\rangle$, Then, at time $t$, quantum state of the QOMG becomes $|\psi(t)\rangle=U(t)|\alpha, \beta\rangle$, which has the following form
\begin{equation}\label{B8}
|\psi(t)\rangle=\sum_n e^{-\frac{|\alpha|^2}{2}} \frac{\alpha^n}{\sqrt{n!}} e^{i E_n^2 \tau(t)} e^{-i E_n \tilde{\beta}(t)}|n\rangle |\mu_n(t)\rangle.
\end{equation}
where $\mu_n(t)=\beta e^{-i \tilde{\omega}_m t}-E_n\left(1-e^{-i \tilde{\omega}_{m} t}\right)$. Eq.~(\ref{B8}) is Eq.~(\ref{eq:15}) in the main text.

	%
	
	%
	
\end{document}